\pdfoutput=1
\documentclass[9pt]{article}

\usepackage{mathtools}
\usepackage{amsfonts}
\usepackage{bbm}
\usepackage{bm}
\usepackage{siunitx}
\usepackage{booktabs}
\usepackage{booktabs}
\usepackage[a4paper, total={6in, 9in}]{geometry}
\usepackage{color}
\newcommand{\new}[1]{#1}

\usepackage{fancyhdr} 
\title{From flagellar undulations to collective motion:\\ predicting the dynamics of sperm suspensions}

\author{Simon F. Schoeller and Eric E. Keaveny\footnote{\texttt{{e.keaveny@imperial.ac.uk}}} \vspace{.5cm} \\ \textit{\small Department of Mathematics}\\ \textit{\small Imperial College London}\\ \textit{\small South Kensington Campus}\\\textit{\small London SW7 2AZ, UK} } 
\date{{\small \today}}

\begin{document}
\maketitle

\begin{abstract}
Swimming cells and microorganisms are as diverse in their collective dynamics as they are in their individual shapes and propulsion mechanisms.  Even for sperm cells, which have a stereotyped shape consisting of a cell body connected to a flexible flagellum, a wide range of collective dynamics is observed spanning from the formation of tightly packed groups to the display of larger-scale, turbulence-like motion.  Using a detailed mathematical model that resolves flagellum dynamics, we perform simulations of sperm suspensions containing up to 1000 cells and explore the connection between individual and collective dynamics.  We find that depending on the level of variation in individual dynamics from one swimmer to another, the sperm exhibit either a strong tendency to aggregate, or the suspension exhibits large-scale swirling.  Hydrodynamic interactions govern the formation and evolution of both states.  In addition, a quantitative analysis of the states reveals that the flows generated at the time-scale of flagellum undulations contribute significantly to the overall energy in the surrounding fluid, highlighting the importance of resolving these flows. 
\end{abstract}


Swimming cells and microorganisms encompass the entire range of cell types and exhibit a great variety of cell geometries and swimming strategies used to move in viscosity dominated environments \cite{Purcell1977, Lauga2009}.  Some organisms, such as Volvox colonies, can be nearly spherical and utilize for propulsion thousands of relatively short flexible flagella distributed along their surface. Others, like sperm cells, have a single, long flagellum that propagates bending waves, leading to a large time-periodic shape change of the cell.  Though stereotyped, the propagated waveforms, frequencies, and head shapes of sperm cells vary appreciably across different species \cite{Pitnick2009}.  

There is not only great variation amongst individual cells, but also in the patterns that they form and in how populations are organized.  This can range from algal plume formation in bioconvection \cite{Pedley1992, Hill2005} to turbulence-like swirling in bacterial baths \cite{Wu2000, Dombrowski2004, Koch2011, Wensink2012}.  Great variation is exhibited by sperm suspensions of different species that have been observed to form vortices near planar boundaries \cite{Riedel2005}, aggregate into coherent sperm trains \cite{Moore2002}, or exhibit turbulence-like swirling as first documented nearly \num{70} years ago by Lord Rothschild \cite{Rothschild1948, Rothschild1949, Creppy2015,Creppy2016}.  What is not clear, however, is how the individual differences in sperm cells, such as geometry, flagellum length, waveform and flexibility, give rise to the differences in the way sperm populations organize themselves.  Mathematical modeling and simulation provide a route to explore the connection between individual and collective dynamics where experiments might otherwise be very challenging.

In the past \numrange{10}{15} years, there has been much work on the development of mathematical models \cite{Koch2011,Aranson2007,AditiSimha2002, Hernandez2005, Saintillan2008, Hohenegger2010,Baskaran2009, Saintillan2007, Saintillan2011, Marchetti2013} of microorganism suspensions, especially to understand the turbulence-like state found in bacterial baths.  To deal with large numbers of cells, the models rely on a reduced description of the swimmers, often treating them as simple, rigid objects (e.g., rods, dumbbells, spheres, ellipsoids) that interact through steady, dipolar flows and steric repulsion.  These models have been effective in reproducing the large-scale motion of the population and relating its formation to the sign of the coefficient of dipolar flow induced by individuals.  

While such models now broadly shape our thinking about suspensions of swimming microorganisms, they reduce the diversity of the microscopic world into a handful of parameters and a limited number of degrees of freedom. They ignore the complexity and time-dependence of cell shapes, as well as the time-dependent and beyond-dipolar features of the flows generated by the shape changes.  Indeed, many of these details have been included in models of individual or small collections of swimmers.  In the case of sperm cells, the addition of flagellar motion leads to hydrodynamically-induced attraction, synchronization, and phase-locking of planar, flagellar waveforms \cite{Fauci1990, Elfring2009, Olson2015}.  In simulations of \numrange{50}{100} swimmers \cite{Yang2008, Yang2010} these effects were found to induce aggregation and clustering of cells.  This tendency to aggregate may aid sperm in forming coherent groups such as sperm trains, however, it remains unclear how a turbulence-like state could then be reached.

In this paper, we perform detailed simulations of up to \num{1000} interacting sperm, resolving the coupled flagellar dynamics along with the flows generated at the undulation time-scale and at sub-flagellum length scales.  We report that variation in the individual dynamics across the population, here controlled by the undulation frequency, can suppress the tendency to aggregate and instead lead to a density dependent turbulence-like state with hydrodynamics still playing a key, but now different role.  Additionally, an analysis of the clustered and turbulence-like states reveals the strong influence of flagellar undulation on the quantities used to explore large-scale motion in swimmer suspensions.  These results suggest that only minor variations in sperm behavior across species are needed to produce very distinct collective dynamics.  

%\begin{align}\label{eqn_stokes}
%    \eta \Delta \bm{u} -\nabla p + \bm{f} = 0,\quad \nabla \cdot \bm{u} = 0
%\end{align}
%with the viscosity $\eta$, the velocity $\bm{u}$, force density due to immersed swimmers $\bm{f}$, and %the pressure $p$.
%++++++++++++++++++++++++++++++++++++++++++++++++++++++++++++++++++++++++++++++++++++++++++++++++++++++++++++
%++++++++++++++++++++++++++++++++++++++++++++++++++++++++++++++++++++++++++++++++++++++++++++++++++++++++++++
\subsection*{Mathematical model}
Our swimmer model is closely related to several models \cite{Fauci1988, Smith2009b, Olson2011, Montenegro2012, Simons2015} employed to describe the dynamics of sperm cells in a viscous fluid and extends directly from a model previously used to capture undulatory locomotion through structured media \cite{Majmudar2012}.  We briefly summarize it here in the context of a single swimmer and provide a more detailed description in the SI.  

Our swimmers consist of two elements, the flagellum and the cell head.  The flagellum is treated as an inextensible, yet flexible beam of length $l$ and bending modulus $K_B$, while an oblate spheroid of semi-major and minor axes $a$ and $b$, respectively, represents the cell head.  The overall swimmer length is taken to be $d = 2.1a + l$, \new{where the distance $0.1a$ accounts for a linkage between the flagellum and the head.}  \new{The flagellum is parametrized by arclength $s \in [0,l]$ such that the position of a point along the flagellum is $\mathbf{Y}(s)$ and the unit tangent at that point is $\bm{\hat{t}}(s) = d\mathbf{Y}/ds$.  The flagellum is driven internally by the moments per unit length, \new{$\bm{\tau}^D$}, that arise due to the preferred curvature,
\begin{align}
    \kappa(s,t) &= K_0\sin\left(k s -\omega t + \varphi \right) \cdot \begin{cases}
2(l - s) / l,\quad &s > l/2 \\
1,\quad & s\le l/2 , \label{eqn_kappa}\end{cases}
\end{align} 
where $\textcolor{black}{\omega  = 2\pi/T}$ is the undulation frequency, $k$ is the wavenumber, $\varphi$ is the phase, and $K_0$ is the amplitude.  \new{The linear decay in the preferred curvature amplitude near the free-end is introduced to better reproduce observed flagellar waveforms.}  Allowing the flagellum to also be subject to external applied forces,  $\bm{f}$, and torques, $\bm{\tau}$, per unit length due to viscous stresses, the force and moment balances along the flagellum are 
\begin{align} \label{eqn_beam1}
\frac{d \bm{\Lambda}}{d s} + \bm{f} &= 0\\ 
\frac{d \bm{M}}{d s} + \bm{\tau}^D + \bm{\hat{t}}\times \bm{\Lambda}+\bm{\tau} &= 0. \label{eqn_beam2}
\end{align}
where $\bm{\Lambda}$ is the tension and $\bm{M} = K_B \bm{\hat{t}}\times d\bm{\hat{t}}/ds$ is the bending moment.}
\new{At one end $(s=0)$, the flagellum is attached via a clamped-end condition to the cell head, while the other end $(s = l)$ remains free.}    

The continuous beam equations, \new{(\ref{eqn_beam1}) and (\ref{eqn_beam2})}, are discretized to obtain force and torque balances on $N_\text{flag}$ segments of the flagellum, as well as the cell head (see SI for details). For the $n$-th segment of the flagellum, the balances are
\begin{align}\label{eqn_force_balance}
	\bm{F}^{C}_n  + \bm{F}^{H}_n &= 0, \\
	\bm{T}^{B}_n + \bm{T}^{C}_n  +  \bm{T}^{D}_n + \bm{T}^{H}_n &= 0,\label{eqn_torque_balance}
\end{align}
where $\bm{T}^{B}_n$ are torques arising due to bending, $\bm{F}^{C}_n$ and $\bm{T}^{C}_n$ are constraint forces and torques associated with tension, $\bm{T}^{D}_n$ are the torques due to the preferred curvature, and $\bm{F}^{H}_n$ and $\bm{T}^{H}_n$ are the hydrodynamic forces and torques on the segment.  \new{The force and torque balances yield a low Reynolds number mobility problem for the translational and angular motion of the segments that is identical to that for a collection of rigid particles.}  We solve this mobility problem using a regularized multipole approach for Stokes flows known as the \emph{force-coupling method} (FCM) \cite{Maxey2001, Lomholt2003, Liu2009, Majmudar2012, Maxey2017}.  In the limit of large $N_\text{flag}$, FCM provides an approximation of the hydrodynamics consistent with a regularized version of slender body theory \cite{Batchelor1970, Johnson1979, Lighthill1996, Tornberg2004, Cortez2012, Smith2009b, Montenegro2012}.  \new{After solving the mobility problem for the velocities and angular velocities of the flagellum segments and cell head, the differential equations for their positions and orientations are integrated in time using a second-order backwards differentiation scheme.  Broyden's method is then used to solve the resulting system of equations for the updated positions and orientations, and the Lagrange multipliers associated with the forces and torques arising from the inextensibility constraint.}

We choose the parameters in our simulations to reproduce a cell geometry, flagellar waveform, and swimming speed close to that reported for ram sperm \cite{Denehy1975, Creppy2015} and in the range of other mammalian sperm \cite{Rikmenspoel1965, Woolley1977}.  The head size $a$ is such that $a/l = 0.0470$ and $b = a/3$.  The wavenumber is $ k = 2\pi/l$, and the curvature amplitude is $K_0 = 12.76/l$.  Finally, we set the dimensionless parameter known as the \emph{Sperm number}, which provides a measure of the ratio of the viscous to elastic forces \cite{Lowe2003, Keaveny2008, Lauga2009, Majmudar2012, Delmotte2015}, to $\text{Sp} = \left(4 \pi \omega \eta/K_B \right)^{1/4} l = 12.0$, where $\eta$ is the dynamic viscosity.  With these parameters, 23.7 undulation periods are required for the sperm to swim its flagellum length.

We perform three dimensional simulations in domains of size $L\times L \times L_z$ corresponding to 2D periodic thin films with finite thickness $L_z = 0.277d$.  \new{The corresponding fluid flow is resolved on a grid of size $N_x \times N_y \times N_z$}. Free surface conditions at $z=0$ and $z=L_z$ are established by restricting the motion of the swimmers to the mid-plane $z = L_z/2$ and applying periodic boundary conditions in all three directions.  This particular choice in boundary conditions and film thickness corresponds to those commonly employed in experiments on swimming microorganisms \cite{Wu2000, Aranson2007, Guasto2010, Kurtuldu2011}.  %$L_z = 0.3041\cdot l$

\new{When simulating multiple swimmers, a short-ranged repulsive force between nearly touching segments is included to capture steric repulsion between swimmers (see SI). The strength of this force at contact is determined by the parameter $F_S$. This and all simulation parameters, including their values in simulation units, are summarized in Table~1.}
\new{
\begin{table}[h]
\centering %\textcolor{blue}%
{
\caption{Simulation parameters	}
\begin{tabular}{@{}llc@{}}
\toprule
\textbf{Parameter} & \textbf{Description} & \textbf{Value in simulation units} \\ \midrule
    $\eta$           &      Viscosity                             &     $1$                  \\
    $\text{Sp}$      &     Sperm number                           &     $12$                 \\
    $K_0$            &     Curvature amplitude                    &     $0.2$                \\
    $K_B$            &     Bending modulus                        &     $1800$                \\
    $b$              &     Segment radius and head height         &     $1$                  \\
    $a$              &     Head half axis                         &      $3$                  \\
    $d$              &     Swimmer length                         &   $70.1$                   \\
    $F_S/d$          &     Steric reference force	              &   $15\pi$              \\
    $N_{x,y}$        &     Grid dimensions (in-plane)             &     $ \{2048,3072\}$ \\
    $N_z$            &     Grid dimension (normal)                &     $64$                   \\
    $L$              &     Linear domain size                     &     $931.3$\\
    $L_z$            &     Domain height                          &     $19.40$  \\
    $N_\text{flag}$  &     Number of flagellum segments per swimmer        &     $29$                   \\
    $N$    &     Number of swimmers                     &     $ \{100,\dots,1000\}$     \\
    $N_t / T$    &     Time-steps per undulation      &     $300$                  \\
    $\sigma_\omega/\omega$  &     Relative std. dev. of frequencies      &    $ \{0, 0.2 \}$      \\
    \bottomrule
\end{tabular}}
\label{tab_parameters_simulation}
\end{table}
}

\begin{figure}[h]
    \centering
    \includegraphics[width=\linewidth]{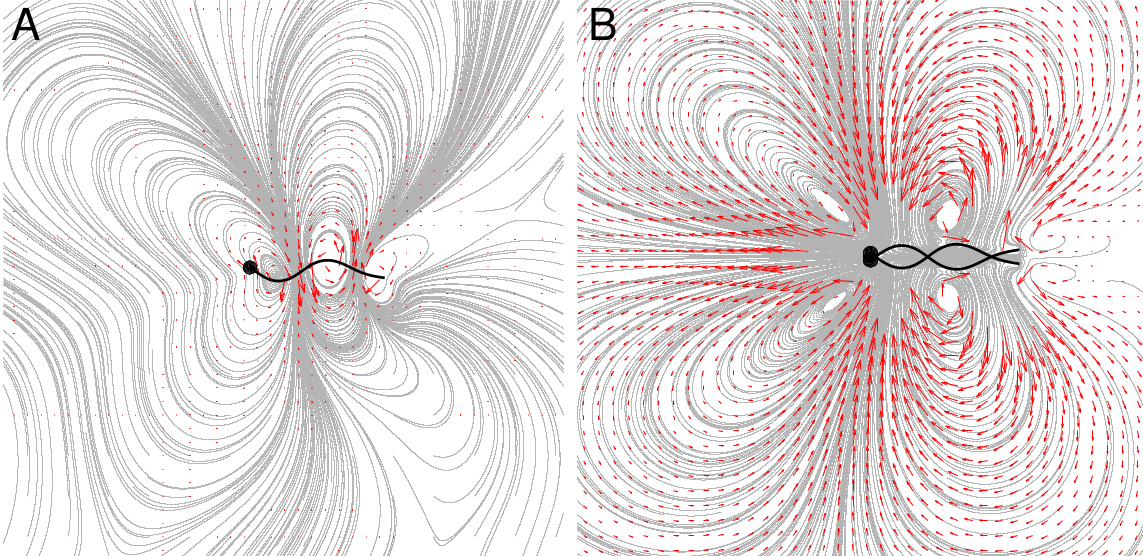}
    \caption{Fluid flow induced by an individual swimmer. \textbf{A} Instantaneous streamlines (grey) and fluid velocity field (red arrows) at $z = L_z/2$ produced by a single swimmer at one instant in time. \textbf{B} Period-averaged streamlines and fluid velocity field in a co-moving frame at $z = L_z/2$. Two snapshots of the swimmer are also shown.}
    \label{fig:panel_streamlines}
\end{figure}

\subsection*{Time-dependent and higher-order features of the swimmer flow field}
Examining the flow field produced by a single swimmer and the force-moments associated with this flow, we see clear differences between their instantaneous and period-averaged values.  The flow at the mid-plane of the thin film and the swimmer at an instant in time are shown in Fig. \ref{fig:panel_streamlines}A, while Fig. \ref{fig:panel_streamlines}B shows the flow averaged over one undulation period.  These flow fields are comparable to those given by other computational studies of individual swimmers  propelled by an elastic filament \cite{Fauci1988, Yang2008, Yang2010, Gaffney2011, Olson2015, Ishimoto2017}.  While the period-averaged flow closely resembles the dipolar flow generated by a so-called pusher, we find that the instantaneous flow field at any point in time is markedly different, containing features of higher-order forces singularities.  
\begin{figure}[h]
    \centering
    \includegraphics[width=\linewidth]{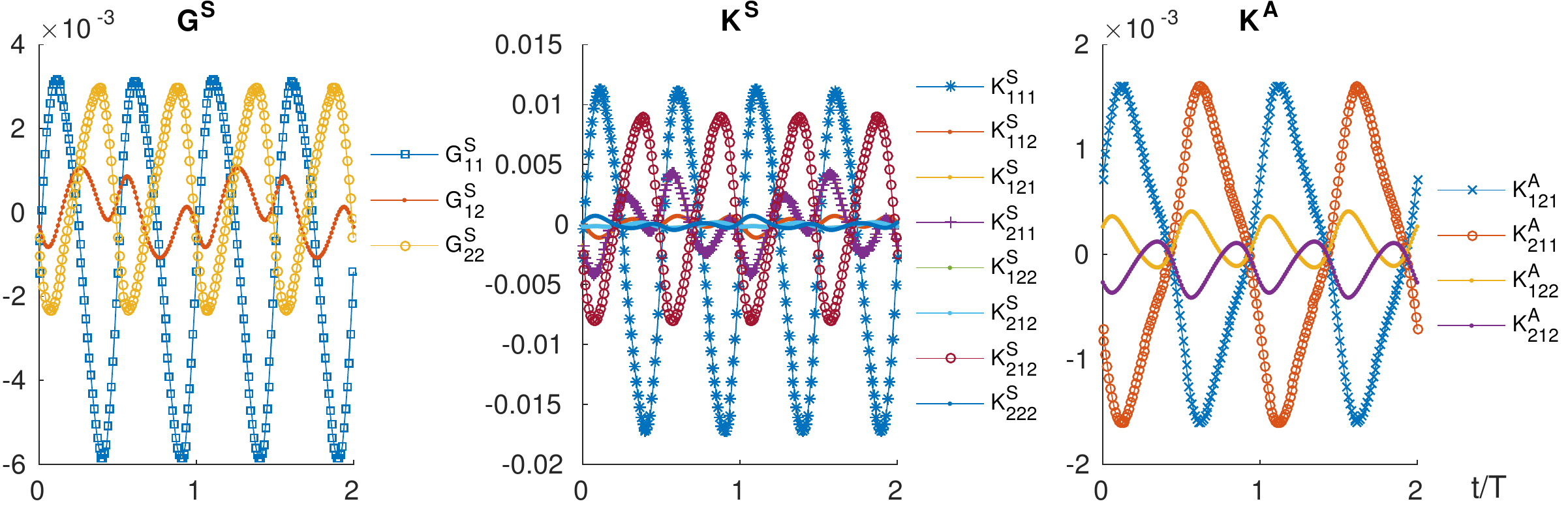}
    \caption{The symmetric, trace-less dipole, $G_{ij}$, and quadrupole, $K_{ijk}^{S,A}$, coefficients as a function of time.  The values are reported in the swimmer frame where the origin is the swimmer's center of mass and $\hat{e}_1$ is the average swimming direction.  Components that are zero at all times are excluded.  Dipole entries are in units of $F_0d$, while the quadrupole entries have units of $F_0d^2$.}
    \label{fig:panel_moments}
\end{figure}%
To analyze this further, we compute (see SI and \cite{Keaveny2008}) the in-plane dipole and quadrupole force moments for the swimmer as functions of time.  The symmetric, trace-less dipole tensor $G$, the symmetric (in the first two indices), trace-less quadrupole tensor $K^S$, and the anti-symmetric quadrupole tensor $K^A$, are plotted over two undulation periods in Fig. \ref{fig:panel_moments}.  The time periodic nature of the force moments, especially their change in sign over the course of a period, are consistent with experimental measurements \cite{Guasto2010, Ishimoto2017} on flagellate microswimmers.  The anti-symmetric dipole moment is always zero, as there is no net-torque on the swimmer.  From these time series, we can extract the non-vanishing time-averages, the largest of which are the dipole coefficients $\langle (G^S_{11}, G^S_{22})\rangle = (-1.050,  0.432)\times10^{-3} F_0 d$, and the quadrupole coefficients $\langle (K^S_{111}, K^S_{221}) \rangle = (-2.321, 0.822)\times10^{-3} F_0 d^2$, where $F_0$ is the in-plane drag on the cell head moving at the average swimming velocity.  We note that for both the dipole and quadrupole these average values are much smaller than the maximum values attained during an undulation period.  Quadrupole components ($K^S_{111}$ and $K^S_{221}$) are indeed expected to contribute substantially to the fluid flow around spermatozoa \cite{Smith2009c, Montenegro2012, Montenegro2012} and have also been linked to swimmer velocity correlations in bacterial suspensions \cite{Liao2007}.  The maximum value of the dipole coefficient in our model is more than a factor of $5.5$ greater than its mean value, while for the quadrupole this increases to a factor of $7$.  Additionally, the relatively large values of the quadrupole coefficients produce strong flows that, although they decay faster than the dipolar flow, are non-negligible for significant distances from the swimmer.  We estimate that the instantaneous quadrupolar flow dominates the dipolar flow up to separations of approximately $3d$.  

From this, a picture emerges that not only are the time-dependent aspects of the dipolar flow much stronger than the average flow, but that higher-order contributions are non-negligible when time dependence is included.  Both of these features can impact the swimmer-swimmer hydrodynamic interactions, especially in semi-dilute to dense suspensions.  As we will see in the proceeding sections, these features, linked to the dynamics at the level of individual cells, do indeed propagate to larger scales and can significantly affect the observed coherent structures and the processes by which they form.
%++++++++++++++++++++++++++++++++++++++++++++++++++++++++++++++++++++++++++++++++++++++++++++++++++++++++++++

%++++++++++++++++++++++++++++++++++++++++++++++++++++++++++++++++++++++++++++++++++++++++++++++++++++++++++++

\begin{figure}[h]
    \centering
    \includegraphics[width=\linewidth]{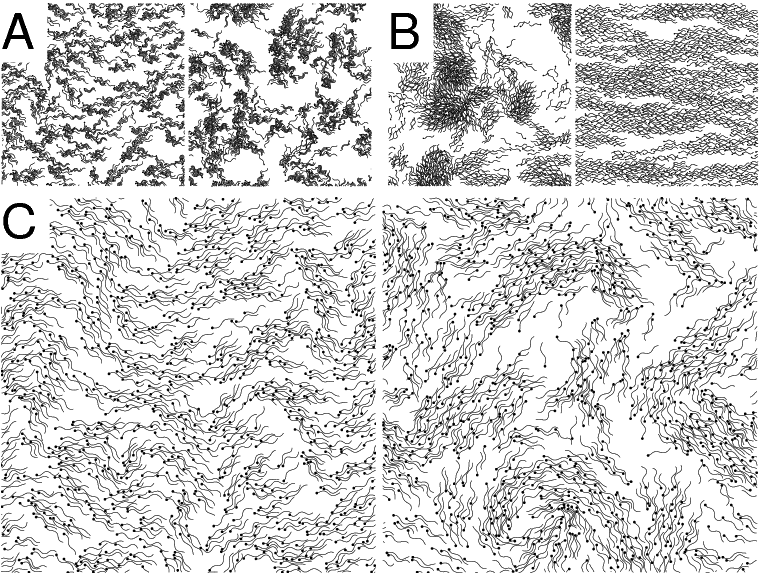}
    \caption{Evolution of suspensions of 1000 swimmers. \textbf{A} Snapshots of a system with fixed undulation frequency taken after 30 (left) and 198 (right) periods. \textbf{B} Snapshots after 340 undulation periods from two stochastically-varying frequency RFT simulations starting from isotropic (left) and polar (right) initial configurations. \textbf{C} Snapshots of a simulation with randomly varying frequencies starting from a polar configuration at 60 (left) and 198 (right) undulations.}
    \label{fig:panel_collective}
\end{figure}%
%++++++++++++++++++++++++++++++++++++++++++++++++++++++++++++++++++++++++++++++++++++++++++++++++++++++++++++
\subsection*{Suspensions with a monodisperse distribution of undulation frequencies form clusters}
Using the methods described in the previous section, we simulate the dynamics of a suspension of $N = 1000$ swimmers with $N_\text{flag} = 29$ for \num{200} undulation periods.  In this simulation, all swimmers are driven by a prescribed preferred curvature, Eq. \ref{eqn_kappa}, with the same undulation frequency, $\omega = K_B (\text{Sp}/l)^4/(4\pi\eta)$,  and wave-number $k = 2\pi / l$.  Each swimmer has a different phase, $\varphi$, drawn randomly from a uniform distribution.  The swimmers are initially distributed uniformly in the domain and are aligned to swim in the $-x$-direction.  Swimmer-swimmer hydrodynamic interactions are incorporated by solving the full mobility problem that couples the motion of all swimmer segments and heads, while steric interactions are included through short-ranged, pairwise, repulsive forces between all flagellum segments and cell heads.  The effective area fraction is $\nu = N d^2/(4L^2) = 1.42$. Based on the film thickness, the effective volume fraction, as used in \cite{Saintillan2007}, is approximately \num{2.56}.  For suspensions of pusher dipoles at this effective volume fraction, one would expect rapid decay of the initial polar order followed by the onset of large-scale, fluctuating motion \cite{Saintillan2007,Saintillan2008}.
%\eric{We should give a 1 sentence summary of what we'd expect based on reduced models, such as for this density rod models (continuum and discrete) predict that pushers are unstable.  Is this true?}

Fig. \ref{fig:panel_collective}A shows the suspension after \num{30} and \num{198} undulation. periods.  The corresponding video showing the complete evolution is provided as ESM.  While the suspension quickly evolves from the initial polar state, the dynamics are very different from the long-wavelength bending instabilities predicted by both simulations of rigid, steady pushers \cite{Hernandez2005, Saintillan2007} and kinetic or continuum models \cite{AditiSimha2002, Saintillan2008, Hohenegger2010,Baskaran2009, Marchetti2013}.  Instead, we find that the swimmers have a strong tendency to aggregate and form clusters within which flagella are aligned. A quantitative analysis of this cluster growth is included in the SI.  
Once small clusters have formed, they remain intact and can attract other clusters in their vicinity.  As two attracting clusters approach each other, they either rotate to swim in the same direction and merge to form a bigger, synchronized cluster, or instead swim in opposing directions and move apart at an enhanced relative velocity.  Earlier simulations of sperm \cite{Yang2008, Yang2010} showed similar aggregation dynamics in two dimensional suspensions, but here we see that aggregation also dominates in larger, denser suspensions where long-range hydrodynamics could be expected to produce large-scale swirls and jets.  %\eric{We can move the comments about the swirling in Creppy's experiments to the beginning of the next section where we show how to recover them.}

\begin{figure}[h]
    \centering
    \includegraphics[width=\linewidth]{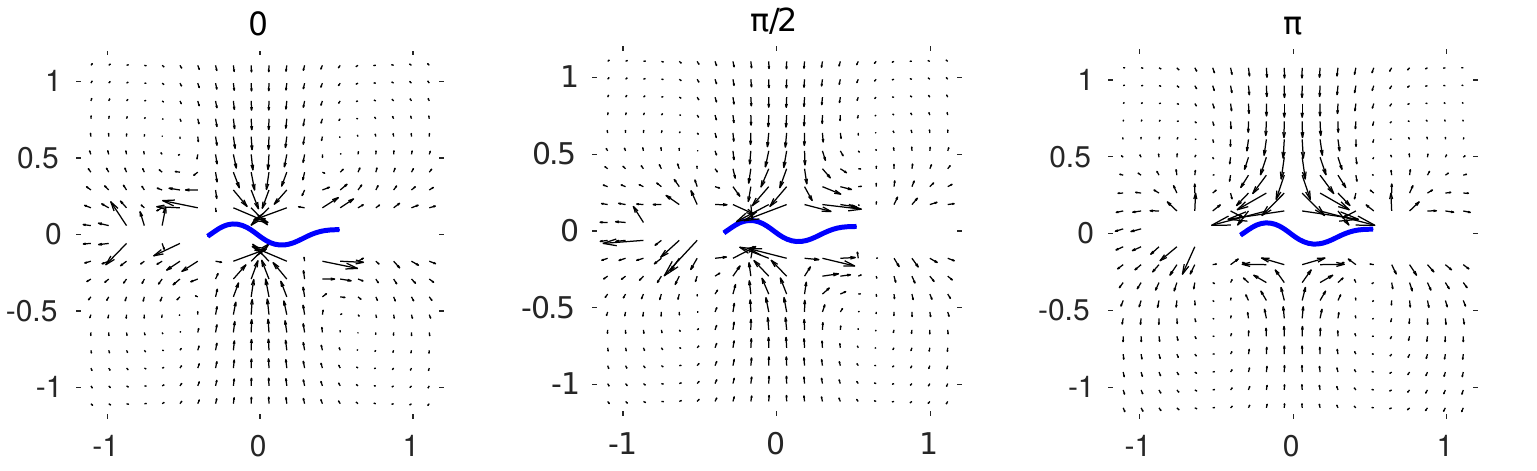}
    \caption{Relative center of mass displacements (in units of swimmer length) over four periods of undulation for two parallel swimmers with relative phase shifts $\Delta\varphi = 0,\pi/2$ and $\pi$.  For clarity, the cell heads attached to the left ends of the flagella are omitted.}
    \label{fig:panel_attraction}
\end{figure}%

The aggregation process is driven by the hydrodynamic attraction and phase-locking behavior that has been found previously with models of undulatory swimmers \cite{Elfring2009, Yang2008, Yang2010, Llopis2013, Simons2015, Olson2015}.  We measure this for our system by examining in detail the pairwise interactions between two swimmers moving in the $-x$-direction with phase differences $\Delta\varphi = 0, \pi/2,$ and $\pi$.  The vector fields in Fig. \ref{fig:panel_attraction} show the displacement over four undulation periods of one swimmer centered at $(x,y)$ relative to another placed at the origin.  When the swimmers are far apart, all three vector fields resemble the displacement field associated with pusher dipoles.  At smaller separations, however, the fields differ and the influence of the phase difference can be readily seen.  We see that there is clear attraction, if the swimmers are in phase.  When the phase difference is $\pi/2$, the swimmers attract and simultaneously move relative to each other as to align the crests of the flagellar waves.  If they are out of phase, the swimmers again move relative to one another attempting to align wave crests, however, the relative shift is much larger.  The large displacements observed near the swimmer's ends are due to the steric forces between the head of one swimmer and the flagellum of the other.

%++++++++++++++++++++++++++++++++++++++++++++++++++++++++++++++++++++++++++++++++++++++++++++++++++++++++++++
\subsection*{Suspensions with stochastically varying frequencies reach a turbulence-like state}
In the previous section, we saw that flagellum undulations can produce flows that lead to hydrodynamically-induced aggregation and cluster formation.  While this provides a mechanism for the formation of coherent groups, such as wood mouse sperm trains \cite{Moore2002}, the patterns differ markedly from the large-scale swirling found in experiments on ram sperm suspensions \cite{Creppy2015,Creppy2016}.  In real samples, sperm cells have distributions of waveforms, undulation frequencies or geometries.  Accordingly, the parameters appearing in our model should not necessarily be the same for all individuals and should be chosen to accurately describe variations from individual to individual across the population.  

We explore how variations in individual dynamics across the population impact large-scale suspension dynamics by introducing stochastic modulations of the undulation frequency into the model, allowing flagellar waves to differ from individual to individual and to vary over time \cite{Rikmenspoel1965, Friedrich2010}.  As we solve for the flagellum dynamics, changing the frequency also leads to changes in the waveform.  The individual frequencies are drawn from a log-normal distribution after time intervals of length $T = 2\pi/\omega$ as described in the SI.  We set the mean frequency to $\omega$ and, in view of matching experimental data \cite{Denehy1975, Friedrich2010}, set the standard deviation $\sigma_\omega$ to be $\omega/5$.  

We repeat our simulations of \num{1000} swimmers, but now allowing for stochastic variations in the frequencies.  Snapshots of the suspension evolution are shown in Fig. \ref{fig:panel_collective}C and corresponding videos can be found as part of the ESM.  The stochastic variations in the waveforms suppress the ability of neighboring flagella to synchronize.  Rather than forming clusters, the swimmers now form waves, swirls and vortices, as seen in concentrated, quasi-two-dimensional ram sperm suspensions \cite{Creppy2015}.  By running an equivalent simulation starting from an isotropic initial configuration, it was verified that the qualitative features of the fully non-linear state are independent of the initial conditions. \new{We also note that a sufficiently large $\sigma_\omega$ is needed to successfully suppress aggregation and observe large-scale motion of the suspension.  Simulations of smaller ensembles (see SI) show that lower values of $\sigma_\omega$ still exhibit cluster formation (e.g., $\sigma_\omega = \omega/20$ or  $\sigma_\omega = \omega/50$).  In these cases, stochastic frequency changes are rarely large enough to break the synchronization of adjacent flagella.}

While hydrodynamic interactions no longer lead to aggregation, they still play a strong role in the evolution of the suspension.  We perform simulations where we have removed the hydrodynamic interactions by solving the mobility problem using resistive force-theory (RFT) instead of FCM.  As describe in the SI, the drag coefficients appearing in the RFT are chosen as to closely match the swimming speeds and flagellar waves obtained with the full FCM model.  Fig. \ref{fig:panel_collective}B shows snapshots from two different RFT simulations with $\nu = 1.42$ and $N = 1000$ where the initial swimmer directions are either all aligned or distributed isotropically.  The corresponding videos are provided as part of the ESM. The time-evolution of these suspensions is very different from those with hydrodynamic interactions.  Most strikingly, for the initially polar suspension, long bending-waves are entirely absent and at long times we observe “laning” states similar to those found in simulations of self-propelled rods without hydrodynamic interactions \cite{Yang2010, Wensink2012b, Wensink2012}.  When the initial state is isotropic, laning is is not observed (Fig. \ref{fig:panel_collective}B), indicating further differences between the RFT and full simulations.
\new{Additionally, in thin films, the hydrodynamic interactions between the swimmers are longer ranged than they would be in bulk.  As a result, we find that as we increase the film thickness, though the instability of the polar state takes longer to develop, we still observe (see SI) the eventual onset of large-scale motion qualitatively similar to that found in the thinner film cases.}

\subsection*{Order parameters exhibit density dependence}
The polar and nematic order of the suspension is dependent on the effective area fraction, $\nu$.  We perform a series of smaller stochastic simulations ($\sigma_\omega = \omega/5$) ranging from \numrange{170}{500} swimmers run to $t=400 T$ and compare the order parameters
\begin{align}
    S_1 (t) &= \left\langle -\hat{x}\cdot\hat{n}^n(t)\right\rangle_n,\\
    S_2 (t) &= \left\langle 2(\hat{x}\cdot\hat{n}^n(t))^2 -1\right\rangle_n,
\end{align}
which measure the global alignment with the initial swimming direction ($S_1$) and the global nematic order with respect to the initial direction ($S_2$) \cite{Saintillan2007}.  Here, $\hat{n}^n$ is the $n$-th swimmer's mean orientation and $\langle \cdot \rangle_n$ denotes the average over all swimmers.  Videos of the simulations are provided as ESM.

\begin{figure}
    \centering
    \includegraphics[width=\linewidth]{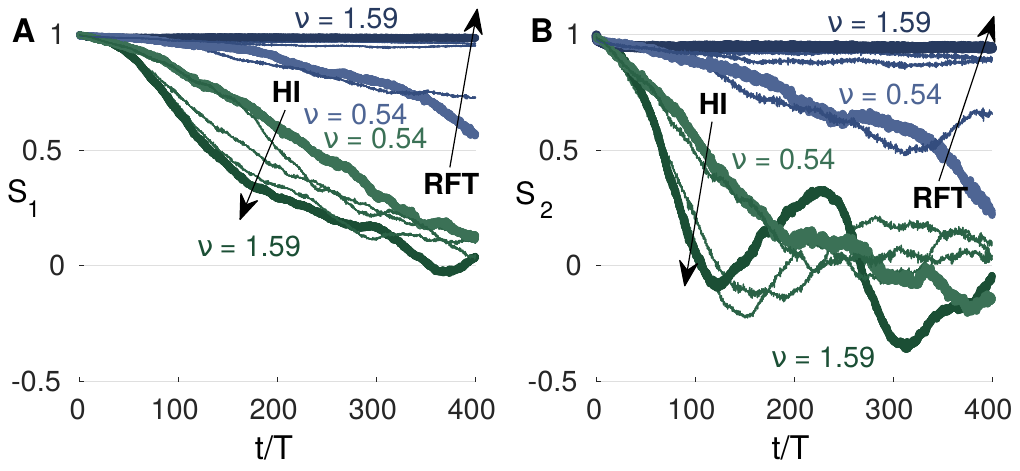}
    \caption{Evolution of the global order parameters $S_1$ (\textbf{A}) and $S_2$ (\textbf{B}) over $400$ periods for initially aligned suspensions with $\sigma_\omega = \omega/5$, $L = 8.86d$, and $\nu = 0.54 - 1.59$.  Simulations are performed with full hydrodynamic interactions (HI, green) and with steric interactions only (RFT, blue), which show the opposite trend with $\nu$.}
    \label{fig:order_parameters_spectra}
\end{figure} 

We find that the directional order, given by $S_1$, decreases with time and the decay rate increases with swimmer density (Fig. \ref{fig:order_parameters_spectra}A).  The time-scale of the decay is comparable to that found for rod models \cite{Saintillan2007,Saintillan2011}.  The nematic order, given by $S_2$, initially decreases with time. However, for higher densities, we see large fluctuations emerge after the initial decay (Fig. \ref{fig:order_parameters_spectra}B).  The appearance of these fluctuations during the relatively gentle decay of $S_1$ reflects periodic folding and rotation of large patches of aligned swimmers.  Fig. \ref{fig:order_parameters_spectra} also shows $S_1$ and $S_2$ for initially polar RFT simulations with the same effective area fractions.  With hydrodynamic interactions removed, we find that higher density suspensions retain their alignment for longer times due to enhanced local caging.  

%. In these RFT simulations, global alignment decays slower, the denser the system, presumably due to enhanced local caging between neighbors. In the RFT instances with the highest area fraction, $S_{\{1,2\}}$ are constant up to small fluctuations. 
%The deviations from unidirectional motion are larger in instances with lower area fraction, in which swimmers are less likely to be caged between their nieghbors. 

%The instances with the lowest swimmer density are most similar to simulations including hydrodynamics at equal density for long times.  Due to the large fluctuations of $S_2$ for the highest density cases including hydrodynamics, it is plausible that a fully non-linear state is reached.
%Due to the slower decay of the global order parameter, all of the RFT simulations only include transient states, whereas 
%The second half of the hydrodynamic simulations reach states that could be considered approximations samples from a stationary ensemble representing a steady macroscopic state.

%++++++++++++++++++++++++++++++++++++++++++++++++++++++++++++++++++++++++++++++++++++++++++++++++++++++++++++
\subsection*{Flagellar undulations contribute significantly to the energy and velocity spectra}
\begin{figure}
    \centering
    \includegraphics[width = \linewidth]{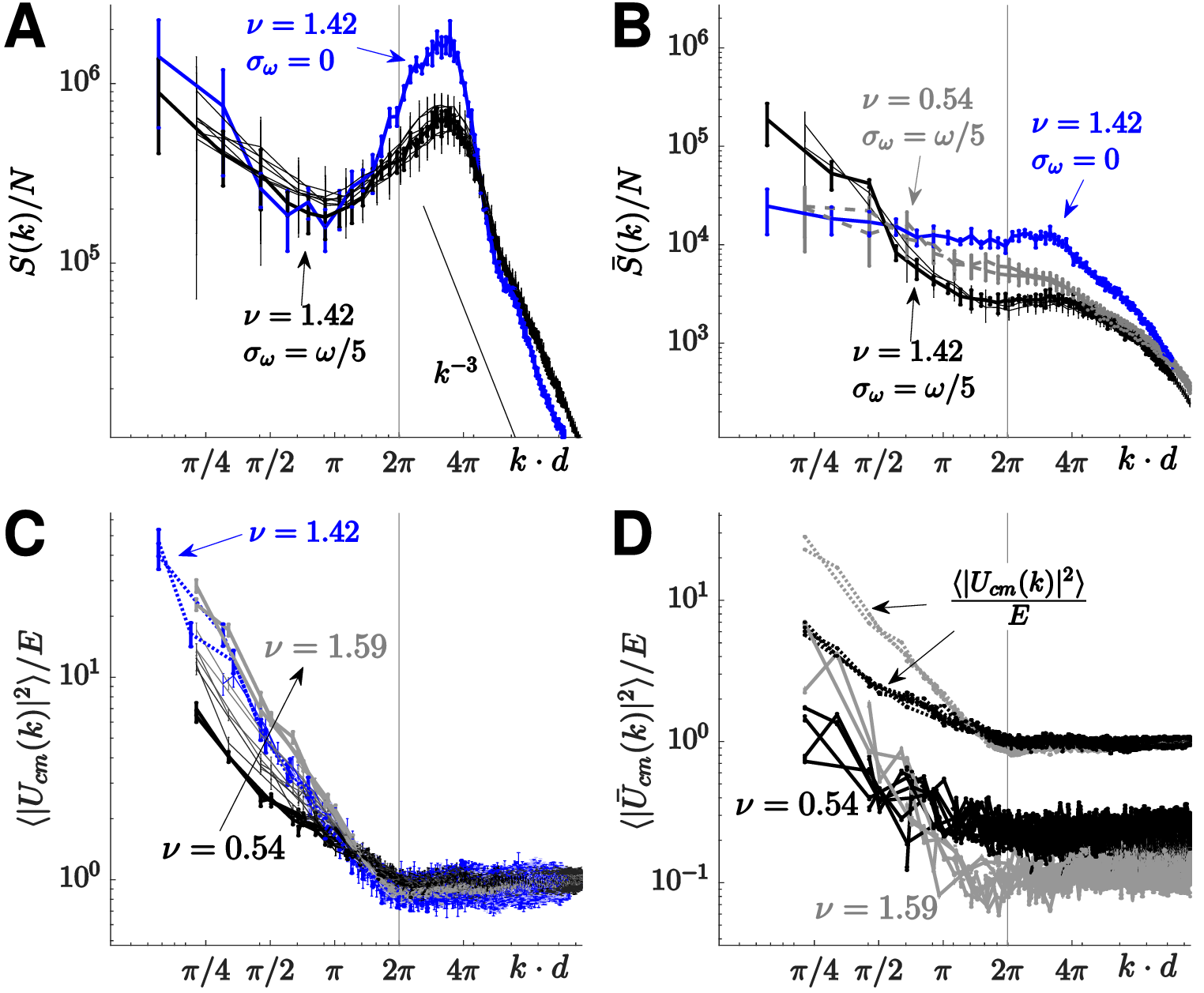}
    \caption{Fluid energy and center of mass velocity spectra for suspensions with effective area fractions $\nu = 0.54 - 1.59$ in domains of size $L = 8.86 d$ and $L = 13.29 d$ with both fixed ($\sigma_\omega = 0$) and stochastically ($\sigma_\omega = \omega/5$) varying frequencies.  For the cases where $L = 8.86 d$ the averaging is performed over times between $200T$ and $400T$, while for $L = 13.29 d$, the interval is $150T$ and $200T$. 
    \textbf{A} Fluid energy spectrum $S(k)/N$. The error bars indicate the sample standard deviation. \textbf{B} Fluid energy spectrum $\bar{S}(k)/N$ of the filtered velocity field from which the short time-scales have been removed.  \textbf{C} Center of mass velocity spectrum for all stochastic simulations.  The data are normalized by the total energy, $E = \sum_m \langle |\bm{U}^{(m)}_{cm}|^2\rangle_t$, for each case.  The error bars indicate one standard error of the mean. \textbf{D} Center of mass velocity spectrum of the filtered center of mass velocity fields for the lowest and highest $\nu$. The spectra for the respective unfiltered fields from C are included for comparison.}
    \label{fig:Ucm_spectra}
\end{figure}

To further quantify the large-scale motion, we compute the fluid energy spectrum,
\begin{align}
    S(k) \equiv \left\langle k\left\langle \left|\hat{\bm{u}}(\bm{k},t)\right|^2\right\rangle_{\|\bm{k}\| = k} \right\rangle_t,
\label{eq:Sofk}
\end{align}
where $\hat{\bm{u}}(\bm{k},t)$ is the spatial Fourier transform of the fluid velocity and $\langle \cdot \rangle_t$ denotes time-averaging while $\langle\cdot\rangle_{\|\bm{k}\| = k}$ signifies averaging over all wave vectors of magnitude $k$.  Fig. \ref{fig:Ucm_spectra}A shows $S(k)/N$ from simulations with both fixed and stochastically varying frequencies and for different values of $\nu$.  With the exception of the pronounced peak near the undulation wavelength, the spectra are similar to those found in experiments on concentrated sperm suspensions \cite{Creppy2015}.  In particular, we also find that the spectra decay like $k^{-3}$ for large $k$ (Fig. \ref{fig:Ucm_spectra}A).  Given that this power-law decay appears over sub-swimmer length scales, we associate it with random forcing of the fluid by the flagella.  A calculation presented in the SI demonstrates that an identical decay is obtained when the fluid is driven by spatially uncorrelated forcing.  This is different from the $k^{-4}$ decay observed at small scales in suspensions of self-propelled rods \cite{Saintillan2011}.  This decay was attributed to the sharp jump in the forcing along the length of each swimmer.  Thus, the details regarding how the swimmers propel themselves can lead to changes in the spectrum at length scales comparable to the swimmer size.

Further, we find that the fluid energy spectrum is dominated by contributions at the time-scale of flagellar undulations, even at long length scales.  Despite qualitative differences in their dynamics and the quantitative differences in the evolution of the order parameters, $S(k)/N$ is nearly identical for all stochastic simulations, regardless of $\nu$.  The collapse illustrates that $S(k)$ scales with the number of swimmers.  This is in contrast with large increases with $\nu$ seen at low $k$ in simulations of self-propelled rod suspensions \cite{Saintillan2011}.  Surprisingly, the $S(k)/N$ curves for the fixed and stochastically varying frequency simulations also do not differ drastically, despite the striking differences in their suspension evolution.    We expect that this is due to the energy injected at the short time-scales associated with flagellar undulations being large compared with that due to the large-scale motion of the suspension.  To limit the contribution of short time-scales to $S(k)$, we apply a low-pass filter to $\hat{\bm{u}}(\bm{k},t)$ by computing a running time-average of $\bm{u}(\bm{x},t)$ with a window of $8T$ before performing the transform.  The spectra, $\bar{S}(k)/N$, computed using the filtered velocity fields are shown in Fig. \ref{fig:Ucm_spectra}B.  With the short time-scales suppressed, the dependence on $\nu$ becomes visible.  We now observe larger values of $\bar{S}(k)/N$ at low $k$ when the sperm density is high, revealing quantitatively the large-scale fluid motion at the time scale of suspension evolution.  Additionally, the spectrum for the fixed-frequency simulations is now almost flat at small $k$ indicative of the lack of long range correlations.

Following from previous work on bacterial suspensions \cite{Wensink2012, Wensink2012b}, we also examine the spectrum of the center of mass velocity field, $\bm{U}_\text{cm}(\bm{x},t) = \sum_{m=1}^{N} \bm{U}_\text{cm}^{(m)}(t)\,\delta\left(\bm{x} - \bm{X}_\text{cm}^{(m)}(t)\right)$, where $\bm{X}_\text{cm}^{(m)}$ and $\bm{U}_\text{cm}^{(m)}$ are the center of mass position and velocity, respectively, as defined in the SI.  Fig. \ref{fig:Ucm_spectra}C shows the swimmer velocity spectrum, $\langle |\bm{\hat {U}}_{\text{cm}}(k)|^2 \rangle_t$, for different values of $\nu$.  For each $\nu$, the lowest spatial modes (lowest $k$) provide the largest contribution to the total spectrum, with the contribution increasing as $\nu$ increases.  From these maximum values, the spectra then quickly decrease with $k$ to a flat profile at scales below the flagellum length, corresponding to the level of noise generated in constructing the swimmer velocity field.  

Due to flagellar undulations, the swimmers' center of mass velocities oscillate about their mean values at the time-scale of the undulation frequency.  We can again remove contributions of the short time-scale by filtering the swimmers' center of mass velocities using a running time-average with window size $8T$.  Fig. \ref{fig:Ucm_spectra}D shows the spectra $\langle|\widehat{\bar{\bm{U}}}_\text{cm}(k)|^2\rangle_t$ associated with the filtered swimmer velocity field.  Filtering reduces values of the swimmer velocity spectrum for all $k$, indicating contributions at short time-scales at all length scales. For lower values of $\nu$, the spectrum is reduced by an approximately constant factor across all $k$, while for higher values of $\nu$, the reduction is more pronounced at shorter scales.  

\subsection*{Discussion and Conclusions}
In this study, we have used numerical simulation to show that variability in sperm flagellum dynamics across a suspension can lead to large changes in collective dynamics.  We have incorporated these variations through stochastic changes in the actuation frequency that, in turn, lead to changes in waveform and amplitude.  In actual sperm suspensions, variations are also likely to be present in flagellum length and cell geometry, and further, the flagellar waveform is likely to be more complex and fully three-dimensional \cite{Woolley1977, Bukatin2015}.  We expect these features to have a similar, aggregation-limiting effect as the stochastic variation in frequency.  As such, measurements quantifying the distributions of individual sperm properties could not only provide a better classification of the individual cells, but also a better understanding of their collective dynamics.

Our fixed frequency simulations show that flagellar synchronization leads to aggregation and clustering.  Certain properties associated with sperm that are found to self-organize into coherent groups, such as wood mouse sperm, might promote synchronization.  It is known that these sperm typically have flagella that are longer than those of sperm from other species.  By allowing for larger amplitudes, longer flagella may enhance the higher-order, time-dependent flows that we find are responsible for aggregation.  Longer, thinner flagella would also bend and flex with greater ease in response to flows generated by neighboring sperm, further reinforcing the tendency to synchronize and aggregate. \new{Additionally, microtubule sliding driving the undulations may depend on the external load experienced by the flagellum.  This could allow for a non-trivial coupling between the actuation mechanism and the surrounding flow field, which, in turn, may also promote synchronisation of neighbouring flagella and cluster formation.}

The turbulence-like state achieved when there is sufficient variation in the undulation frequency is both quantitatively and qualitatively similar to that observed in suspensions of ram and bull semen.  \new{While longer time simulations would be needed to explore this state in more detail, simulations starting from either polar or isotropic states eventually exhibit similar qualitative dynamics and have the same fluid energy spectrum at the final simulation times.  Both this and the many previous results obtained using reduced models suggest that the observed jets and swirls are features of a final turbulence-like state.}  The presence of this state has been empirically correlated with sample fertility \cite{David2015}.  We have shown that its onset depends on sperm density, and its observation could therefore be serving as an indicator of sperm count, a well-known measure of sample fertility.  

While in the simulations presented here the fluid domain is fully three-dimensional, the motion of the sperm cells is restricted to a plane.  Even though layering in sperm suspensions has been observed in experiments \cite{Creppy2015} and sperm are attracted to move along planar surfaces \cite{Rothschild1963}, the imposed two dimensional motion in our simulations may enhance the role of steric interactions.  For example, in a real sample when two sperm cells collide, it is likely that one will pass over the other and they will both experience some change in their swimming directions.  In our simulations, the sperm cannot pass by each other as easily and the collisions can dramatically affect their swimming directions, leading to near alignment or anti-alignment.  The interactions of self-driven filaments \cite{Llopis2013} and the trajectories of interacting sperm pairs \cite{Simons2015} have recently been explored in simulation.  These studies have shown that perturbations from planarity can lead to more complex trajectories, as well as a dependence on the elasticity of the flagella. 
With further advances in numerical methods for simulating flexible filaments and more accurate models for three-dimensional flagellar waveforms, it will be possible to explore these effects through direct simulation of sperm suspensions. 
\new{Our theoretical model for the elastic flagellum and the FCM approach to the mobility problem carry over to fully three dimensional simulations.  While there will be additional computational costs associated with the increased size of the fluid domain, more limiting is the technical challenge involved in keeping track of the fully three-dimensional deformation of the flagellum and, at the same time, using implicit time integration to handle numerical stiffness.  We are currently developing the numerical methods to address this challenge.}

Additionally, it would be of interest to ascertain suspension dynamics at even larger length scales and for longer timescales than those that may be accessible through direct simulation.  Continuum models incorporating the time dependence of the flows generated by the swimmers have recently been developed \cite{Fuerthauer2013,Leoni2014,Brotto2015} and could possibly be adapted to capture the effects of flagellar undulations seen in our simulations and used to explore nonlinear suspension dynamics at these scales.

Along with variations amongst the cells, collective dynamics are likely to be influenced by more complex features of the environment, such as nearby boundaries that could curve sperm trajectories \cite{Friedrich2010}, or particles or other micro-scale structures in the surrounding fluid that are known to enhance swimming speeds of undulatory swimmers \cite{Majmudar2012}.  Non-Newtonian features of the surrounding fluid, such as viscoelasticity, typically associated with biological fluids have also been been found to promote clustering \cite{Tung2017}.  These, and other important, perhaps more complex effects, such as chemotaxis, could further contribute to the richness and diversity of sperm collective dynamics.  

%\acknow{}
\subsection*{Competing interests}
The authors declare no competing interests.
\subsection*{Authors' Contributions}
S.F.S. and E.E.K. designed research, performed research, analyzed
data, and wrote the paper.
\subsection*{Acknowledgements}
The authors would like to thank Pierre Degond and Demetrios Papageorgiou for helpful discussions.
\subsection*{Funding}
S.F.S. gratefully acknowledges funding by an Imperial College President's PhD scholarship. E.E.K. acknowledges support from EPSRC grant EP/P013651/1. 

%\newpage
\subsection*{Supporting Information}
see below %separate document (supporting.tex)

\newpage 
\section*{Supporting Information}
\subsection*{Swimmer model}
As each swimmer is identical, for clarity, we describe the model for a single swimmer and subsequently describe how it generalizes when considering a suspension.  \new{Recall that a swimmer is composed of two elements, its cell head and its flagellum, where the cell head is treated as a rigid oblate spheroid with semi-major and minor axes $a$ and $b$, respectively.  Using the arc-length parametrization where $\bm{\hat{t}}(s) = d\mathbf{Y}/ds$, the force and moment balance equations are}
\begin{align*}
\frac{d \bm{\Lambda}}{d s} + \bm{f} &= 0\\
\frac{d \bm{M}}{d s} + \bm{\tau}^D + \bm{\hat{t}}\times \bm{\Lambda}+\bm{\tau} &= 0.
\end{align*}
 where $\bm{\Lambda}$ is the tension and $\bm{M} = K_B \bm{\hat{t}}\times d\bm{\hat{t}}/ds$ is the bending moment. \new{The external force per unit length, $\bm{f}$, and torque per unit length, $\bm{\tau}$, arise due to the viscous stresses along the flagellum and, in the case of multiple swimmers, steric interactions between neighboring swimmers.}  The torques $\bm{\tau}^D$ per unit length arise due to the time-dependent preferred curvature. At the free end $(s = l)$, the boundary conditions are such that the tension and moment are zero, while at the attachment point to the cell head, we require a clamped-end condition be satisfied.

To solve these equations numerically, we first discretize the flagellum into $N_\text{flag}$ segments of length $\Delta L$.  The segments have positions $\bm{Y}_n$ and orientations $\bm{\hat{t}}_n$ for $n = 1,\dots, N_\text{flag}$.  The orientations are the discrete representation of the centerline tangents at the segment positions.  Considering the tension and bending moment at the midpoints between adjacent segments and applying central differencing, we obtain the discretized force and torque balances
\begin{align}
&\frac{\bm{\Lambda}_{n+1/2} - \bm{\Lambda}_{n-1/2}}{\Delta L} + \bm{f}_n = 0 \label{eq:disfbal}\\
&\frac{\bm{M}_{n+1/2} - \bm{M}_{n-1/2}}{\Delta L} + \nonumber\\
&\frac{1}{2}\bm{\hat{t}}_n\times (\bm{\Lambda}_{n+1/2} + \bm{\Lambda}_{n-1/2})+\bm{\tau}^D_n + \bm{\tau}_n = 0 \label{eqn:dismbal},
\end{align}
where $\bm{M}_{n+1/2} = (K_B/\Delta L) \bm{\hat{t}}_n \times \bm{\hat{t}}_{n+1}$.  Multiplying through by $\Delta L$ and introducing $\bm{F}_n = \bm{f}_n\Delta L$ and $\bm{T}_n = \bm{\tau}_n\Delta L$ as the total applied force and torque, respectively, on segment $n$, we may write Eqs. [\ref{eq:disfbal}] and [\ref{eqn:dismbal}] as
\begin{align}
	\bm{F}^{C}_n  + \bm{F}_n&= 0, \label{eqn:force_balance} \\
	\bm{T}^{B}_n + \bm{T}^{C}_n  + \bm{T}^{D}_n + \bm{T}_n &= 0,\label{eqn:torque_balance}
\end{align}
where $\bm{F}^{C}_n =  \bm{\Lambda}_{n+1/2} - \bm{\Lambda}_{n-1/2}$, $\bm{T}^{B}_n =  \bm{M}_{n+1/2} - \bm{M}_{n-1/2}$, and $\bm{T}^{C}_n = (\Delta L/2)\bm{\hat{t}}_n\times (\bm{\Lambda}_{n+1/2} + \bm{\Lambda}_{n-1/2})$.  The driving torques, $\bm{T}^D_n$, arising from the preferred curvature are given by
\begin{align*}
\bm{T}^D_n= K_{B}(\kappa(s_n,t) - \kappa(s_{n+1},t)) \hat{z}
\end{align*}
where $s_n = (n - 1/2)\Delta L$ for $n = 1,\dots, N_\text{flag}$, and 
\begin{align*}
    \kappa(s,t) &= K_0\sin\left(k s -\omega t + \varphi \right) \cdot \begin{cases}
2(l - s) / l,\quad &s > l/2 \\
1,\quad & s\le l/2,  \end{cases}
\end{align*}
as well as, $\kappa(s_1,t) = \kappa(s_{N_\text{flag}+1},t) = 0$.

The tension $\bm{\Lambda}_{n+1/2}$ enforces flagellum inextensibility at the level of each segment.  In the discrete setting, they are the Lagrange multipliers associated with the constraints,
\begin{align}
\bm{Y}_{n+1} - \bm{Y}_n + \frac{\Delta L}{2}(\bm{\hat{t}}_n + \bm{\hat{t}}_{n+1}) = 0.\label{eqn:constraints}
\end{align}

In the discretized system, the boundary conditions at the free end are satisfied by taking $\bm{\Lambda}_{N_\text{flag} + 1/2} = \bm{M}_{N_\text{flag} + 1/2} = 0$.  The clamped-end condition on the cell head is recovered by treating the head as another segment of size $2a$ with the bending moment and inextensibility constraint computed with respect to the attachment point rather than the head center.  

Until now, we have discussed the model in the context of a single swimmer.  To allow for $N$ swimmers, we must have $N$ force and moment balances, one pair for each of the flagella.  As the driving torques, tension, and bending moments are internal to each flagellum, the force and moment balances are coupled only through the external forces and torques.  In our simulations, the coupling arises due to steric repulsion between the segments and heads and hydrodynamic interactions.  Accordingly, the total external force on segment $n$ is given by $\bm{F}_{n} = \bm{F}^S_n + \bm{F}^H_n$ where $\bm{F}^S_n$ is the steric force and $\bm{F}^H_n$ is the hydrodynamic force on the segment.  The only external torque on the segment is the viscous torque, $\bm{T}_{n} = \bm{T}^H_n$

The steric force $\bm{F}^S_n$ on segment $n$ is given by the generic repulsive barrier force
\begin{align*}
	\bm{F}^S_n &=-F_S \sum_{m \,\in\, I(n)}  \left(\frac{(\chi R_{nm})^2 - \left|\bm{Y}_n -\bm{Y}_{m}\right|^2}{(\chi R_{nm})^2 - R_{nm}^2}\right)^4 \frac{\left(\bm{Y}_n - \bm{Y}_{m}\right)}{d},
\end{align*}
where the sum runs over $I(n)$, the set of all segments/heads, $m$, with $\left|\bm{Y}_n -\bm{Y}_{m}\right| < \chi R_{nm}$ and $\chi = 1.1$.  The parameter $R_{nm}$ specifies the center-to-center distance for particles $n$ and $m$ at contact and $\chi$ sets the range over which the barrier force acts.   If $n$ and $m$ are both segments, the contact distance is $R_{nm} = 2b$, where as if $n$ and $m$ are both heads, we have $R_{nm} = 2a$.  Lastly, if $n$ and $m$ are a head-segment pair then $R_{nm} = a + b$.  The parameter $F_S$ sets the strength of the repulsion at contact, which in our simulations is $F_S = 9018 K_B/d^2$.

\subsection*{Force-coupling method} The force and torque balances, Eqs. (\ref{eqn:force_balance}) and (\ref{eqn:torque_balance}) respectively, establish a low Reynolds number mobility problem whose solution is the coupled motion of the particles through the fluid.  To solve this mobility problem, we utilize the force-coupling method (FCM) \cite{Maxey2001, Lomholt2003, Liu2009}, which, for the sake of clarity, we describe here in the case of spherical particles of equal radius, however, it readily extends to spheroidal particles \cite{Liu2009}, such as our cell heads, and allows for polydispersity in particle sizes.

In FCM forces and torques on the particles are projected onto the fluid using a truncated and regularized force multipole expansion. This leads to the incompressible Stokes flow
\begin{align*}
 \eta \nabla^2\bm{u} -\bm{\nabla}p = &\sum_{n} \bm{F}^H_n \Delta_n(\bm{x}) %\\
  - \frac{1}{2}\sum_{n} \bm{T}^H_n \times \bm{\nabla} \Theta_n(\bm{x}) \label{eqn:stokes}\\
\bm{\nabla}\cdot \bm{u} &= 0,
\end{align*}
for pressure field, $p$, and fluid velocity field, $\bm{u}$.  The Stokes flow is generated by the forces and torques each particle $n$ exerts on the fluid, which, in the case of our swimmers, will be $\bm{F}^H_n = -\bm{F}^{C}_n  - \bm{F}^{S}_n$, and $\bm{T}^H_n = -\bm{T}^{B}_n - \bm{T}^{C}_n  -  \bm{T}^{D}_n$.  The force and torques are projected to the fluid via the Gaussian functions, which for particle $n$ are given by
\begin{align*}
\Delta_n(\bm{x}) &=  \left(2\pi \sigma_{\Delta}^2\right)^{-3/2}\exp \left[ -\left|\bm{x} - \bm{Y}_n\right|^2/(2\sigma_{\Delta}^2)\right],\\
\Theta_n(\bm{x}) &= \left(2\pi \sigma_{\Theta}^2\right)^{-3/2}\exp \left[ -\left|\bm{x} - \bm{Y}_n\right|^2/(2\sigma_{\Theta}^2)\right].
\end{align*}
The motion of the particles is obtained by volume averaging the resulting fluid velocity using the same Gaussian functions.  Specifically, the translational and angular velocities are given by 
 \begin{align*}
\bm{V}_n &= \int \bm{u}\,\Delta_n(\bm{x}) \mathop{d^3\bm{x}} \\
\bm{\Omega}_n &= \frac{1}{2}\int (\bm{\nabla} \times \bm{u})\, \Theta_n(\bm{x}) \mathop{d^3\bm{x}}. 
\end{align*}
If the particles have radius $A$, FCM reproduces the correct Stokes drag and viscous torque with $\sigma_{\Delta} = A/\sqrt{\pi}$ and $\sigma_{\Theta}=A/(6\sqrt{\pi})^{1/3}$.  In our simulations, we preserve these ratios between the particle and the Gaussian envelope sizes.  The segments are taken to have radius $b = \Delta L/2.2$, while the semi major-axes of the spheroidal heads are $a = 3b$ and the semi minor-axes are $b$. All simulation parameters are summarized in table 1 and the datasets provided have the same units.

\subsection*{Updating positions and orientations}
As our simulations are quasi-2D in that the motion of the swimmers is restricted to a plane, we have $\bm{\Omega}_n = \Omega_n \bm{\hat{z}}$ for all $n$ and can introduce angle $\theta_n$ such that $\bm{\hat{t}}_n = (\cos \theta_n, \sin \theta_n)$.  Therefore, after computing the $\bm{V}_n$ and $\bm{\Omega}_n$ for each segment and head, the positions and orientations are obtained by integrating in time
 \begin{align}
\frac{d\bm{Y}_n}{dt} &= \bm{V}_n  \\
\frac{d\theta_n}{dt} &= \Omega_n.
\end{align}
subject to the inextensibility constraints given by [\ref{eqn:constraints}].   To do this numerically, we rely on a second-order implicit BDF scheme \cite{Ascher1998} and Broyden's method \cite{Broyden1965} to find the updated positions and orientations, as well as the Lagrange multipliers $\bm{\Lambda}_{n+1/2}$.  

\subsection*{Swimmer center of mass and director}
We define the center of mass of the $m$-th swimmer as
\begin{align*}
    \bm{X}^{(m)}_\text{cm} &= \frac{1}{a^2+ N_\text{flag} b^2} \left(a^2\bm{Y}^{(m)}_0 + b^2\sum_{n=1}^{N_\text{flag}} \bm{Y}^{(m)}_n \right).
\end{align*}
Here, $\bm{Y}^{(m)}_0$ is the position of the $m$-th swimmer's head, and $\bm{Y}^{(m)}_k$ is that of its $k$-the segment.
The center of mass velocity is then given by 
\begin{align*}
    \bm{U}^{(m)}_\text{cm} &= \frac{1}{a^2+ N_\text{flag} b^2} \left(a^2\bm{V}^{(m)}_0 + b^2\sum_{n=1}^{N_\text{flag}} \bm{V}^{(m)}_n \right).
\end{align*}
The director of swimmer $m$ is given by
\begin{align*}
    \hat{{n}}^{(m)} &= -  \frac{1}{N_\text{flag} + 1}\sum_{n=0}^{N_\text{flag}} \bm{\hat{{t}}}^{(m)}_n.
\end{align*}

\subsection*{Force multipole expansion for a single swimmer}
%In the following, when looking at an arbitrary or individual swimmer, the first index is suppressed and we simply write $\bm{X}^n$ instead of $\bm{X}^{(n,1)}$, etc. Where applicable, the summation convention is used.
The singular force density for a single swimmer is given by
\begin{align*}
    f_i(\bm{x}) = \sum_n F_i^n \delta(\bm{x} - \bm{Y}_n) + \sum_n G_{ij}^n \partial_j  \delta(\bm{x} - \bm{Y}_n),
\end{align*}
where the asymmetric dipole is related to the torque through $G_{ij} = \epsilon_{ijk}T_{k}/2$.  We have suppressed the superscript index, since we are considering only one swimmer.  Expanding this force density about the swimmer's center of mass, $\bm{X}_\text{cm}$, and defining the position of the center of mass relative to the $n$-th segment/head as $\bm{R}^n = \bm{X}_\text{cm} - \bm{Y}_n$, we have 
%\begin{widetext}
\begin{align*}
    f_i(\bm{x}) &= \\ &= \underbrace{\left(\sum_n F_i^n\right)}_{= 0} \delta(\bm{x} - \bm{X}_\text{cm}) \,+\\
    &+\left(\sum_n G_{ij}^n + F_i^n R^n_j \right)\partial_j  \delta(\bm{x} - \bm{X}_\text{cm}) \,+\\
    &+ \left(\sum_n G_{ij}^n R^n_k + \frac{1}{2} F_i^n R^n_j R^n_k  \right) \partial_k \partial_j  \delta(\bm{x} - \bm{X}_\text{cm})\, +\\ 
    &+ \text{ higher order terms}.
\end{align*}
%\end{widetext} 
The first term is zero because there is no external force on the swimmer.  The remaining two terms are the force dipole and quadrupole, 
\begin{align*}
    G_{ij}  &\equiv \sum_n G_{ij}^n + F_i^nR^n_j \\
    K_{ijk} &\equiv \sum_n G_{ij}^n R^n_k + \frac{1}{2} F_i^n R^n_jR^n_k ,
\end{align*}
respectively, which drive the flow observed in the limit $\left| \bm{x} - \bm{X}_\text{cm} \right| \gg d$.  We compute the symmetric, trace-less and anti-symmetric parts of these tensors, given by
\begin{align*}
    G_{ij}^S &= \frac{1}{2}(G_{ij} + G_{ji}) - \frac{1}{3} G_{kk}\delta_{ij}\\
    G_{ij}^A &= \frac{1}{2}(G_{ij} - G_{ji})\\
    K_{ijk}^S&= \frac{1}{2}(K_{ijk} + K_{jik}) - \frac{1}{3} K_{mmk}\delta_{ij} \\
    K_{ijk}^A&= \frac{1}{2}(K_{ijk} - K_{jik})
\end{align*}
as functions of time. These are shown in Fig.~2 in the main text, with the exception of $G^A(t)$ which is zero for all time as there is no net-torque acting on the swimmers.
\subsection*{Stochastic variation of the undulation frequency}
We introduce stochastic changes of the undulation frequencies to explore how variability within a population, as well as the fluctuations of individual frequencies over time can impact collective dynamics.  After every fixed average period $T = 2\pi/\omega$, a new undulation frequency, $\omega_B$, is randomly assigned to a swimmer to replace its previous value, $\omega_A$.  New frequencies are drawn from a log-normal distribution with parameters ($\mu_\text{ln}, \sigma_\text{ln}$) given by
	\begin{align*}
		\mu_\text{ln}    &= \log\left(\frac{\omega^2}{\sqrt{\omega^2 + \sigma_\omega^2}}\right)\\
		\sigma_\text{ln} &= \sqrt{\log\left(\delta_\omega^2 + 1\right)},
	\end{align*}
where the mean frequency is $\omega$ and the standard deviation of the distribution is $\sigma_\omega= \delta_\omega \omega$.
To enforce continuity in time for each swimmer's preferred curvature $\kappa(s,t)$, the phase must also be updated from $\varphi_A$ to $\varphi_B$.  Specifically, at the time $t_{AB}$ when the frequency changes, we require
\begin{align*}
    ks - \omega_A t_{AB} + \varphi_A - (ks - \omega_B t_{AB} + \varphi_B) \in 2\pi\mathbb{Z}.
\end{align*}
This can be fulfilled by taking
\begin{align*}
    \varphi_B = 2\pi \left(\frac{(\omega_B - \omega_A)t_{AB} + \varphi_A}{2\pi} - \left \lfloor{\frac{(\omega_B - \omega_A)t_{AB} + \varphi_A}{2\pi}}\right \rfloor\right),
\end{align*}
for each individual swimmer.

We note that an alternative approach to introducing stochasticity is to assign a random frequency to each swimmer at the beginning of the simulation and keep it fixed for all time.  This was adopted previously in \cite{Yang2008}.  We found that on time-scales that are relatively short compared to the evolution of the suspension this leads to qualitatively similar results as those obtained by allowing the frequency to vary over time.  Over longer times, however, we expect, based on our monodisperse frequency simulations, that keeping the frequencies fixed could potentially lead to preferential clustering of swimmers with similar frequencies and hence structurally different dynamics. 

\subsection*{Identifying and counting swimmer clusters}
\begin{figure}
    \centering
    \includegraphics[width=.5\linewidth]{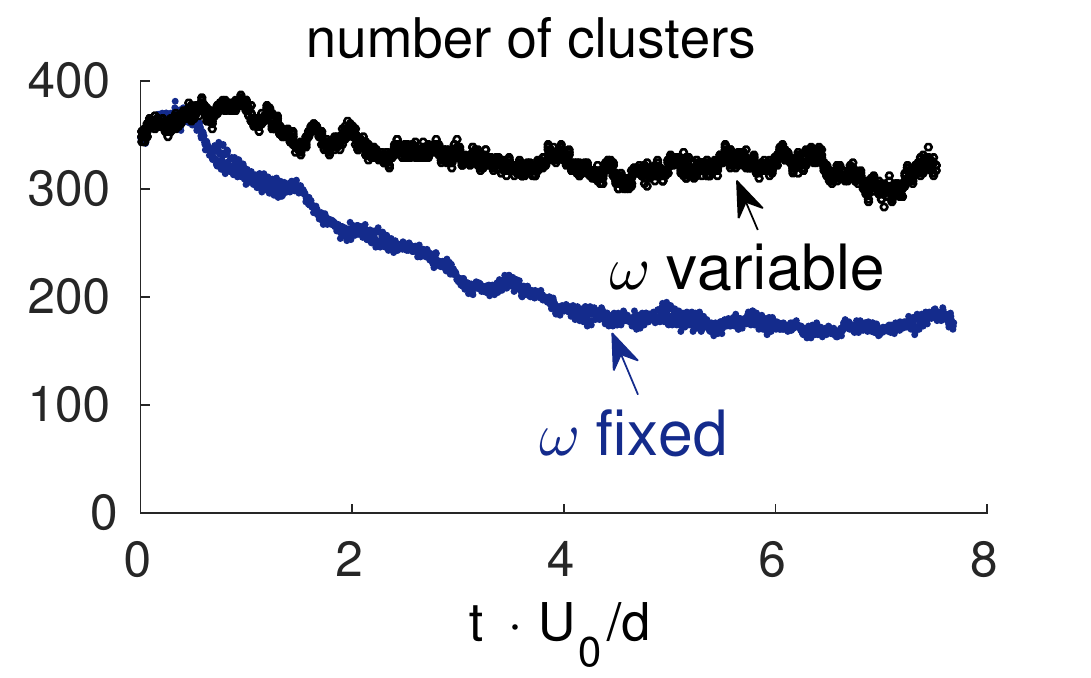}
    \caption{Number of clusters over time in simulations of 1000 swimmers starting from a polar configuration. Time is normalized by the average swimming velocity $U_0$ and the swimmer length $d$.}
    \label{fig:clusters}
\end{figure}
To effectively identify swimmer clusters, we must account for relative swimmer orientations in addition to relative distances.  Based on this, swimmer $i$ is taken to belong to a cluster if the modified distance 
\begin{align*}
    d^\beta(i,j) &\equiv \left|\bm{X}_\text{cm}^{(i)} - \bm{X}_\text{cm}^{(j)}\right| + \beta \tan\left(\frac{1}{2}\arccos\left(\hat{{n}}^{(i)}\cdot \hat{{n}}^{(j)}\right)\right) \\
    &= \left|\bm{X}_\text{cm}^{(i)} - \bm{X}_\text{cm}^{(j)}\right| + \beta \left(\frac{\sqrt{1-\hat{{n}}^{(i)}\cdot \hat{{n}}^{(j)})}}{\sqrt{1+\hat{{n}}^{(i)}\cdot \hat{{n}}^{(j)}}}\right)
\end{align*}
with at least one other swimmer $j$ in the cluster is $d^\beta(i,j) < d/3$.  The parameter $\beta$ controls the influence of particle alignment and is chosen to be $\beta/d = \frac{1}{6}\frac{1}{\tan(\pi/8)} = (\sqrt{2}-1)/6$.  With this choice of $\beta$, swimmers with orientations differing by an angle of $\pi/4$ must be half as far apart as aligned swimmers to still be considered clustered.  Using this notion of distance, we employ the hierarchical cluster algorithm in \textsc{Matlab} for cluster identification.  The number of clusters over time for the two simulations with $1000$ individual swimmers are shown in Fig.~\ref{fig:clusters}.  Individual swimmers are counted as clusters of size one.  We see that the number of clusters for the fixed frequency simulation is much lower than that for the simulation where the undulation frequency is allowed to vary stochastically.  Additionally, we see that when stochasticity is introduced, cluster growth is halted very early in the simulation.

\new{\subsection*{Effect of film thickness on suspension dynamics}
As the flows induced by singular force distributions decay more slowly in 2D than in 3D, we expect the hydrodynamic interactions to play a stronger role as $L_z \rightarrow 0$ and, as a result, the instability to exhibit a faster growth rate for thinner films.  Figs. \ref{fig:LzSweep} and \ref{fig:heightdependency} show results from simulations containing 85 swimmers in domains of linear size $L = 4.43d$ and for thicknesses $L_z = 0.277d, 0.554d$, and $1.107d$.  The effective area fraction is $\nu = 1.08$ and the swimmers are initially in a polar state.
\begin{figure}[h!]
    \centering
    \includegraphics[width=.75\linewidth]{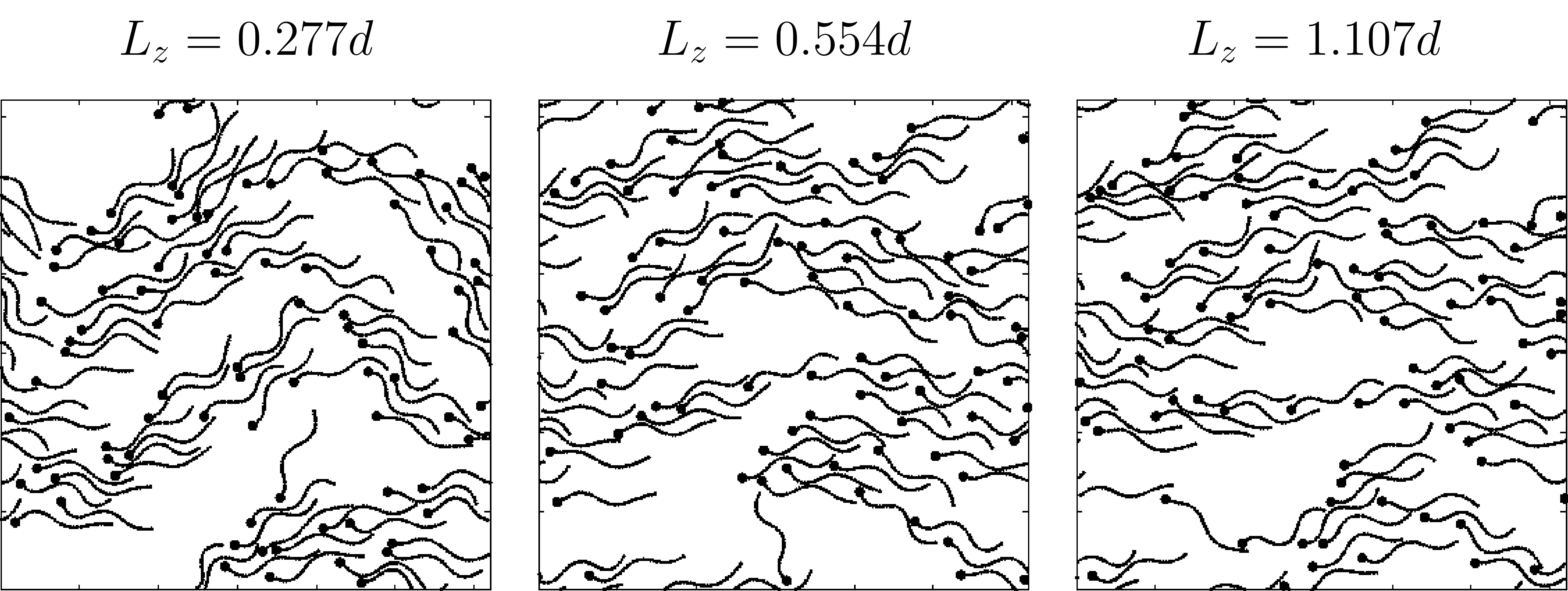}
    \caption{\new{Snapshots at $t= 100T$ of simulations with 85 swimmers all starting from a polar initial state for film thicknesses $L_z = 0.277d, 0.554d, 1.107d$ (left to right)}.}
    \label{fig:LzSweep}
\end{figure}
\begin{figure}[h!]
    \centering
    \includegraphics[width=.45\linewidth]{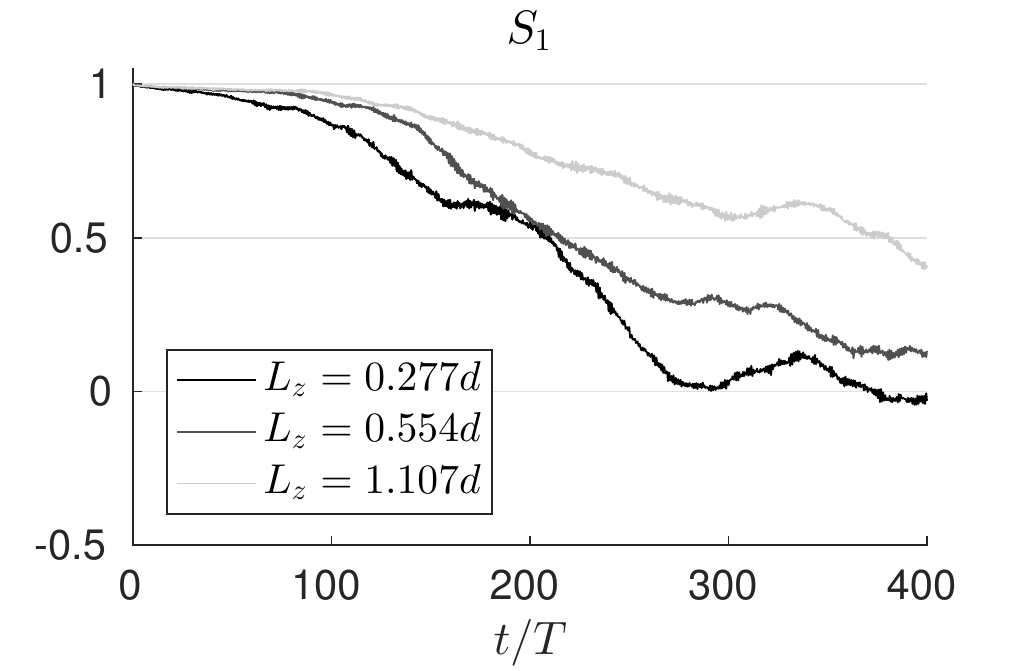}
    \includegraphics[width=.45\linewidth]{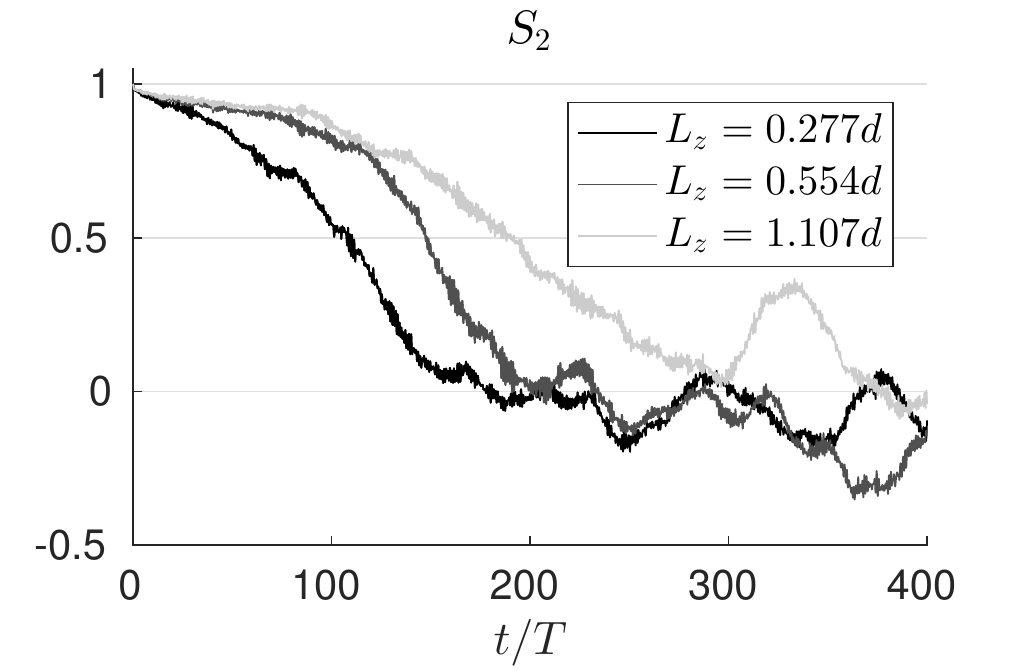}
    \caption{\new{Order parameters as a function of time for the simulations of 85 swimmers in films with thicknesses $L_z = 0.277d, 0.554d,$ and $1.107d$.  The swimmers are initially aligned with the $-x$ direction.}}
    \label{fig:heightdependency}
\end{figure}
In accordance with the slower formation of the bending waves as shown in Fig. \ref{fig:LzSweep}, the slower decay (see Fig. \ref{fig:heightdependency}) of the order parameters $S_1$ and $S_2$ (as defined in the main text) for thicker films indicate that the shorter range of the hydrodynamic interactions lead to slower growth rates of the instability.  The smallest value of $L_z$ used in Fig. \ref{fig:heightdependency} matches that for the larger simulations presented in the main text.  The order parameter decay rate in those simulations, however, is faster still due to the longer modes afforded by the larger system size.}

\new{\subsection*{Effect of undulation frequency variability on aggregation}
As mentioned in the main text, low variability in the swimmers' undulation frequencies does not entirely suppress cluster formation and still allows for neighboring swimmers to synchronize and align their waveforms. 
\begin{figure}[h!]
    \centering
    \includegraphics[width=1.\linewidth]{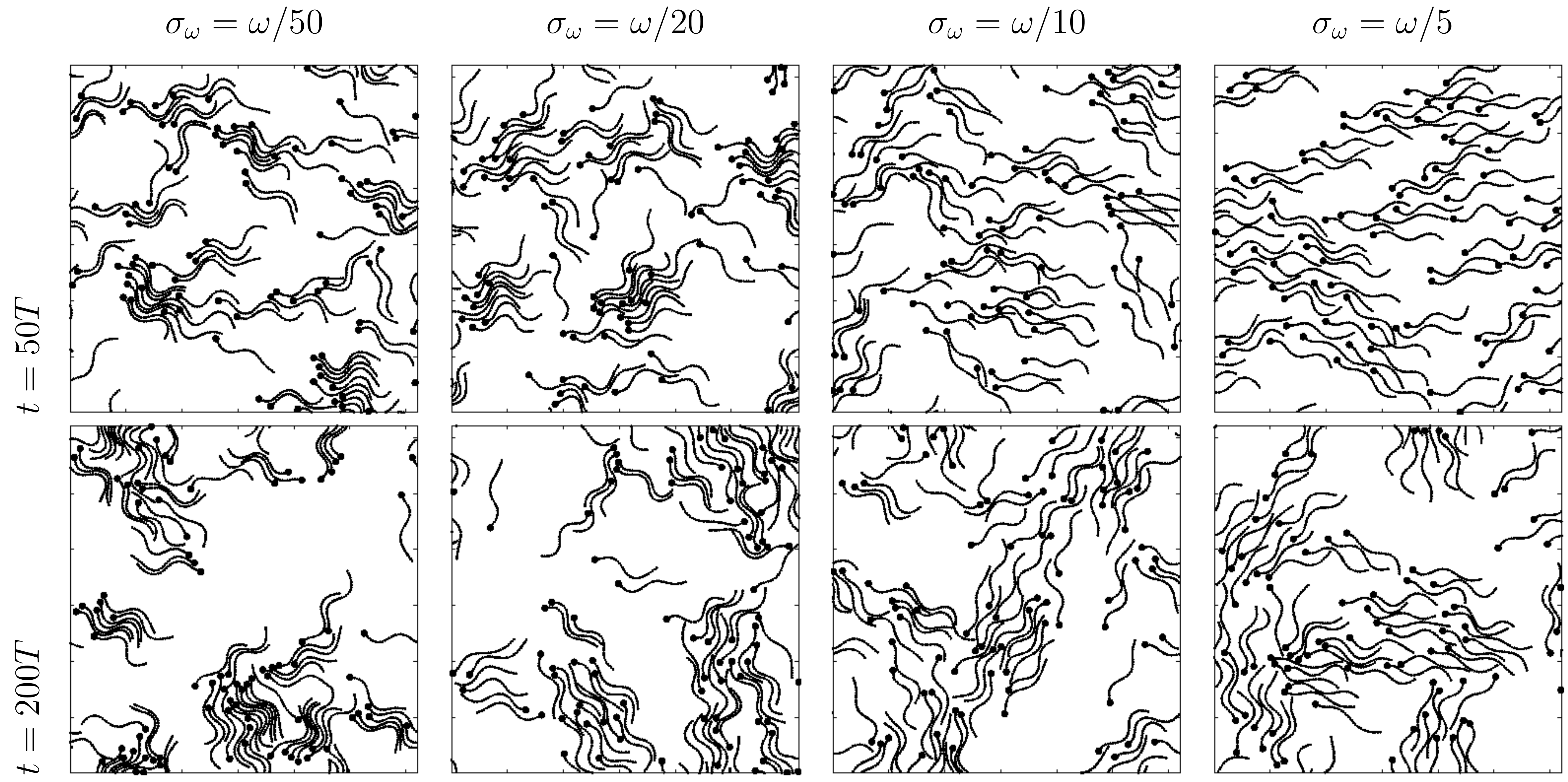}
    \caption{\new{Snapshots at times $t = 50T$ and $t = 200T$ of simulations with $85$ initially aligned swimmer with different frequency distributions.  Each simulation has domain of size $L = 4.43d$ and film thickness $L_z  = 0.277d$. From left to right, the frequency standard deviations are $\sigma_\omega = \omega/50,\, \omega/20,\, \omega/10,\, \omega/5$, while the mean frequency, $\omega = 2\pi/T$, and all other parameters are the same.}}
    \label{fig:SigSweep}
\end{figure}
Snapshots from simulations of a system of size $L = 4.43d$ with $85$ initially aligned swimmers and different values of $\sigma_\omega$ reveal that for small $\sigma_\omega$, cluster formation persists (see Fig. \ref{fig:SigSweep}).  As $\sigma_\omega$ increases, the number of visually identifiable clusters is reduced and we see instead the emergence of a bending wave.}

\subsection*{Tuning the resistive force theory}
\begin{figure}
    \centering
    \includegraphics[width=.5\linewidth]{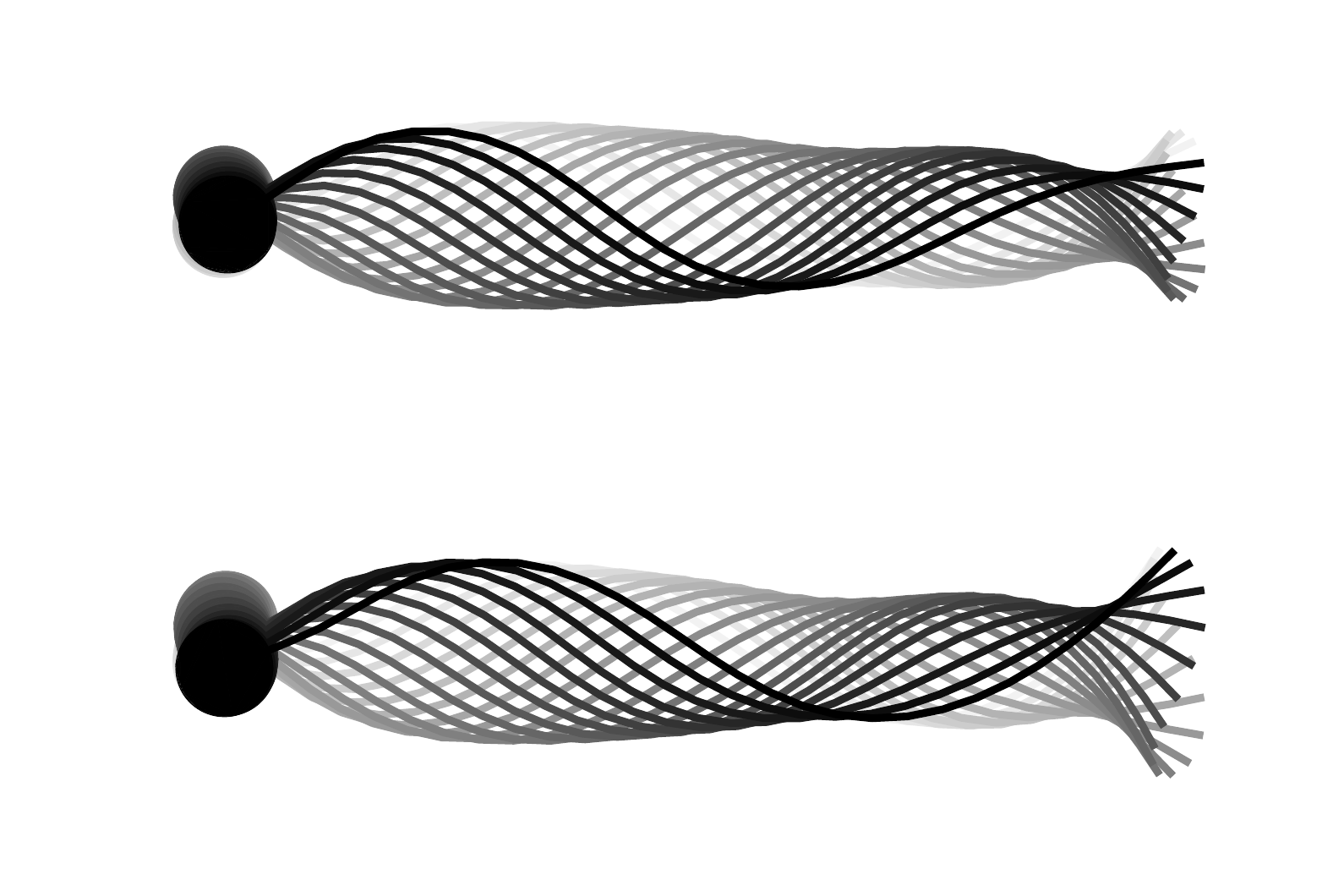}
    \caption{Waveform of an individual swimmer with full hydrodynamics (top) and RFT only (bottom) in the center of mass frame. The swimming direction is to the left and approximately half an undulation period is depicted. The lighter the color of the center-line, the further the snapshot lies in the past. }
    \label{fig:waveform_label}
\end{figure}

To remove the hydrodynamic interactions between the swimmers yet still allow for propulsion via undulation, we solve the mobility problem using a drag-based resistive force theory (RFT) \cite{Lauga2009, Johnson1979} rather than FCM.  With the drag-based model, the velocity and angular velocity of segment $n$ of a flagellum are related to the force and torque on the segment through
\begin{align*}
    \bm{V}_n &= \left( \alpha_\parallel \bm{\hat{{t}}}_n \bm{\hat{{t}}}_n^\top + \alpha_\perp \left(\bm{I} - \bm{\hat{{t}}}_n \bm{\hat{{t}}}_n^\top\right)\right) \bm{F}_n,\\
     \bm{\Omega}_n &= \beta \bm{T}_n,\quad \beta = (8\pi b^3 \eta)^{-1},
\end{align*}
where $\alpha_\parallel$, $ \alpha_\perp$, and $\beta$ are mobility coefficients for the segments.  These mobility coefficients are initially estimated based on FCM simulations of straight filaments and subsequently adjusted to ensure that the swimmer waveform and free swimming velocity are comparable to those given by the simulations with FCM, see Fig. \ref{fig:waveform_label}.  
For the cell heads, we have for all swimmers $m$
\begin{align*}
    \bm{V}^{(0)}_m &= \gamma \bm{F}^{(0)}_m,\\
    \bm{\Omega}^{(0)}_m &= \lambda \bm{T}^{(0)}_m,
\end{align*}
where $\gamma$ and $\lambda$ are mobility coefficients taken from FCM simulations of a single spheroid.  The numerical values used in the simulations are
\begin{align*}
    &\gamma  = 0.0280(\eta b)^{-1}, \quad
    &\lambda = 0.002430\eta^{-1} b^{-3},\\
    &\alpha_\perp =  0.1100(\eta b)^{-1},\quad
    &\alpha_\parallel = 0.1700(\eta b)^{-1}.
\end{align*}

\subsection*{Energy density spectrum of a randomly forced Stokesian fluid}
In this section, we show that a $k^{-3}$ behavior of the fluid energy spectrum can be related to spatially uncorrelated forcing of a 2D fluid.  In the following, the Fourier transform and its inverse are defined as 
\begin{align*}
 \hat{g}(\bm{k}) &= \int_{\mathbb{R}^2} g(\bm{x}) \textrm{e}^{-i\bm{k}\cdot \bm{x}} \mathop{d^2\bm{x}},\\
 g(\bm{x}) &= \frac{1}{4\pi^2}\int_{\mathbb{R}^2} \hat{g}(\bm{k}) \textrm{e}^{i\bm{k}\cdot \bm{x}} \mathop{d^2\bm{k}}.
\end{align*}

We begin by considering a force density, $\bm{f}(\bm{x})$, restricted to a square domain, 
\begin{align*}
	\bm{f}_L(\bm{x}) &= \bm{f}(\bm{x}) \mathop{\mathbbm{1}_{\Omega}}(\bm{x}),\\
\end{align*}
where $\Omega = [-L/2, L/2]\times [-L/2, L/2]$ and $\mathbbm{1}_{\Omega}$ is the indicator function defined on the set $\Omega$.  The Fourier transform of the restricted force density is then
\begin{align*}
\bm{\hat{f}}_L(\bm{k}) &= \int_{\mathbb{R}^2} \bm{f}_L(\bm{x}) \textrm{e}^{-i\bm{k}\cdot \bm{x}} \mathop{d^2\bm{k}}.
\end{align*}
The Fourier transform of the fluid velocity resulting from the restricted force density is given by
\begin{align*}
\bm{\hat{u}}_L(\bm{k}) = \frac{1}{\eta k^2} \left(\bm{I} - \bm{\hat{k}}\bm{\hat{k}}^\top\right)\bm{\hat{f}}_L(\bm{k})
\end{align*}
where $(1/\eta k^2) \left(\bm{I} - \bm{\hat{k}}\bm{\hat{k}}^\top\right)$ is the Fourier representation of the inverse Stokes operator, $k = |\bm{k}|$, and $\bm{\hat{k}} = \bm{k}/k$.  

The fluid energy density spectrum associated with $\bm{\hat{u}}_L(\bm{k})$ is 
\begin{align*}
    S_L(k) = \frac{k}{2\pi L^2}  \int_0^{2\pi} \left\langle|\bm{\hat{u}}_L(\bm{k})|^2\right\rangle \mathop{d\theta},
\end{align*}
where $\theta$ is related to the two-dimensional wave-vector through $\bm{k} = (k \cos \theta, k \sin \theta)$ and $\langle\cdot\rangle$ denotes the ensemble average.  The energy density spectrum of the infinite system will then be given by $S(k) = \lim_{L\rightarrow\infty} S_L(k)$.

As
\begin{align*}
 |\bm{\hat{u}}_L(\bm{k})|^2 &= \bm{\hat{u}}_L(\bm{k})\cdot \bm{\hat{u}}_L(-\bm{k}) \\
 &= \frac{1}{\eta^2 k^4} \left(\bm{I} - \bm{\hat{k}}\bm{\hat{k}}^\top\right) : \left(\bm{\hat{f}}_L(\bm{k})\bm{\hat{f}}_L^\top(-\bm{k})\right),
\end{align*}
the fluid energy density spectrum is directly related to the forcing through
\begin{align*}
 S_L(k) = \frac{1}{2\pi\eta^2 k^3L^2} \int_{0}^{2\pi} \left(\bm{I} - \bm{\hat{k}}\bm{\hat{k}}^\top\right) : \left\langle \bm{\hat{f}}_L(\bm{k})\bm{\hat{f}}_L^\top(-\bm{k})\right\rangle \mathop{d\theta}.
\end{align*}

For $S_L(k) \propto k^{-3}$ to hold, we must have $\left\langle \bm{\hat{f}}_L(\bm{k})\bm{\hat{f}}_L^\top(-\bm{k})\right\rangle$ independent of $k$.  If, in addition, we require statistically independent and identical force components that do not depend on $\theta$, we have that 
\begin{align*} %\label{Fspacecorrelation}
 \left\langle \bm{\hat{f}}_L(\bm{k})\bm{\hat{f}}_L^\top(-\bm{k}) \right\rangle = AL^2 \bm{I}.
\end{align*}
where $A$ is a constant independent of $L$.  The $L^2$ dependence on the system size guarantees that $S_L(k)$ is well-defined and non-zero as $L \rightarrow \infty$ and that the total energy will scale with the system size. 

Assuming that the forcing is a statistically stationary spatial process, we may express the correlations of the force density as 
\begin{align*}
  &\left\langle \bm{f}_L(\bm{x}+\bm{y})\bm{f}_L^\top(\bm{x}) \right\rangle \\
  &= \frac{1}{L^2} \int_{\Omega} \left\langle \bm{f}_L(\bm{x}+\bm{y})\bm{f}_L^\top(\bm{x}) \right\rangle \mathop{d^2\bm{x}}\\
  & = \frac{1}{L^2} \left\langle \int_{\Omega}  \bm{f}_L(\bm{x}+\bm{y})\bm{f}_L^\top(\bm{x}) \mathop{d^2\bm{x}} \right\rangle .
\end{align*}
As the force density is strictly restricted to $\Omega$, we may replace the integral over $\Omega$ with one over all of $\mathbb{R}^2$.  Thus,
\begin{align*}
  &\left\langle \bm{f}_L(\bm{x}+\bm{y})\bm{f}_L^\top(\bm{x}) \right\rangle \\
  &= \frac{1}{L^2} \left\langle \int_{\mathbb{R}^2}  \bm{f}_L(\bm{x}+\bm{y})\bm{f}_L^\top(\bm{x}) \mathop{d^2\bm{x}} \right\rangle,
\end{align*}
which, from the convolution theorem, may be expressed as
\begin{align*}
  &\left\langle \bm{f}_L(\bm{x}+\bm{y})\bm{f}_L^\top(\bm{x}) \right\rangle  \\   &=   \frac{1}{4\pi^2L^2}\int_{\mathbb{R}^2} \left\langle \bm{\hat{f}}_L(\bm{k})\bm{\hat{f}}_L^\top(-\bm{k}) \right\rangle  \textrm{e}^{i\bm{k}\cdot \bm{y}} \mathop{d^2\bm{k}}. \\
\end{align*}
Finally, using the expression for $\left\langle \bm{\hat{f}}_L(\bm{k})\bm{\hat{f}}_L^\top(-\bm{k}) \right\rangle$ given above, we see that 
\begin{align*}
    \left\langle \bm{f}_L(\bm{x}+\bm{y})\bm{f}_L^\top(\bm{x}) \right\rangle 
    &=  \frac{AL^2\bm{I}}{4\pi^2L^2}\int_{\mathbb{R}^2}\textrm{e}^{i\bm{k}\cdot \bm{y}} \mathop{d^2\bm{k}}\\
    &= A\bm{I} \delta(\bm{y})
\end{align*}
for any $L$.  Allowing $L \rightarrow \infty$ then yields
\begin{align*}
  \left\langle \bm{\hat{f}}(\bm{x}+\bm{y})\bm{\hat{f}}^\top(\bm{x}) \right\rangle& = A \bm{I} \delta(\bm{y})
\end{align*}
and we see that a $k^{-3}$ decay in the fluid energy density spectrum is given by uncorrelated spatial forcing of the fluid.

\end{document}